\documentclass[aps,pre,twocolumn,superscriptaddress,superscriptreference]{revtex4-1}
\pdfoutput=1

\usepackage{amsmath,bbold,bm,amssymb,scalerel,mathtools,latexsym}
\usepackage{graphicx}
\usepackage{color}
\usepackage{enumitem}
\usepackage{algorithm,algpseudocode}
\usepackage{multirow}
\usepackage{colortbl,booktabs}
\usepackage{placeins}
\usepackage{slashbox}

\usepackage[colorlinks, citecolor=blue, urlcolor=blue, linkcolor=blue]{hyperref}

\begin{document}

\title{Topological disentanglement of linear polymers under tension}

\author{Michele Caraglio}
\email{michele.caraglio@uibk.ac.at}
\affiliation{Institut f\"{u}r Theoretische Physik, Universit\"{a}t Innsbruck, Technikerstra{\ss}e  21A, A-6020 Innsbruck, Austria}

\author{Boris Marcone}
\affiliation{Istituto Tecnico Economico Tecnologico Statale `L. Einaudi', via Tommaso D’Aquino 8, 36061 Bassano del Grappa, Italy}

\author{Fulvio Baldovin}
\affiliation{Dipartimento di Fisica e Astronomia and Sezione INFN Universit\`a di Padova, Via Marzolo 8, I-35131 Padova, Italy}

\author{Enzo Orlandini}
\affiliation{Dipartimento di Fisica e Astronomia and Sezione INFN Universit\`a di Padova, Via Marzolo 8, I-35131 Padova, Italy}

\author{Attilio L. Stella}
\affiliation{Dipartimento di Fisica e Astronomia and Sezione INFN   Universit\`a di Padova, Via Marzolo 8, I-35131 Padova, Italy}

\date{\today}

\begin{abstract}
We develop a theoretical description of the topological disentanglement occurring when torus knots reach the ends of a semi-flexible polymer under tension.
These include decays into simpler knots and total unknotting.
The minimal number of crossings and the minimal knot contour length are the topological invariants playing a key role in the model.
The crossings behave as particles diffusing along the chain and the application of appropriate boundary conditions at the ends of the chain accounts for the knot disentanglement.
Starting from the number of particles and their positions, suitable rules allow reconstructing the type and location of the knot moving on the chain 
Our theory is extensively benchmarked with corresponding Molecular Dynamics simulations and the results show a remarkable agreement between the simulations and the theoretical predictions of the model.
\end{abstract}

\maketitle

\section{Introduction}

From tying shoelaces to maneuvers in sailing or mountaineering, knots are useful in our everyday lives.  
At microscopic scales, knots have been shown to occur naturally in DNA~\cite{Sogo1999,Rybenkov1993,Shaw1993,Arsuaga2002}, proteins~\cite{Taylor2000,Virnau2006,Mallam2010}, and more generally in long polymer chains~\cite{Orlandini2007}.
Within polymer rings, knots move and fluctuate in size permanently (equilibrium state) but if tied in open chains they do not represent a genuine topological state of the system and can disappear, form again, or change the underlined topological complexity (knot type)~\cite{Orlandini2018,Tubiana2013,Micheletti2014}.
Yet, these ``physical knots", act as long-lived constraints and can, for instance, affect the metric and mechanical properties of the hosting chain, interfere with the elongation processes induced by either confinement~\cite{Metzler2006,Mobius2008,Micheletti2014} or tensile forces~\cite{Caraglio2015,Panagiotou2014}, hinder the ejection dynamics of viral DNAs from their capsids~\cite{Arsuaga2005,Matthews2009,Marenduzzo2009,Marenduzzo2013}, and slow down the translocation process through nanopores~\cite{Suma2015,Plesa2016,Suma2017}.

Nowadays, physical knots can also be artificially formed in biomolecules via optical tweezers~\cite{Arai1999,Bao2003}, compression in nano-channels~\cite{Amin2018}, and by applying elongational flows~\cite{Renner2014,Narsimhan2017,Klotz2017,Soh2018} or electric fields~\cite{Klotz2018,Klotz2020}.
These single-molecule experiments combined with computational studies~\cite{Metzler2006,Mobius2008,Narsimhan2017,Klotz2017,Soh2018,Soh2020,Matthews2010, Vologodskii2006,Huang2007,Trefz2014,DiStefano2014,Narsimhan2016,Soh2019,Xu2020} have paved the way for a better physical understanding of the mobility of the knot along the chain and on the disentanglement process of the hosting polymer.

Perhaps the most controlled setup to study knot untying in fluctuating chains uses tensile forces applied at their ends.  
In this case, for sufficiently large forces, the physical knot can be easily detected and followed during its erratic motion along the chain (contour motion).
Moreover, unknotting often occurs when the knot reaches one of the chain ends, making the comparison among theory, experiments, and simulations easier.  
In this  regime of highly tensioned chains  knots tend to be  localized within the chain 
and their contour motion is essentially a diffusion process~\cite{Matthews2010,Vologodskii2006,Huang2007}.

Much less is known, however, for moderate and small tensions, where physical knots are more difficult to detect and can widely fluctuate in size~\cite{Caraglio2015,Caraglio2019} giving rise to the known weak localization regime in the limit of zero forces~\cite{Marcone2005,Tubiana2013}.
Valuable insight into this problem has been gained by Ben-Naim \textit{et al.}~\cite{BenNaim2001}, who studied knot untying in tension-free granular chains supported by a vibrating plane and flattened in 2D by gravity.
In this work the statistical properties of unknotting are shown to be well described by a simple model of $n_{\mathrm{k}}$ random walks interacting via hard-core exclusion in one spatial dimension (single-file diffusion), with $n_{\mathrm{k}}$ being the number of essential crossings associated to the knot type $\mathrm{k}$.
This analysis, however, is restricted to the simpler case of quasi-2D chains in which the position of the crossings do not depend on the projection.
This is not the case for the full 3D problem where the spatial location of crossings is very elusive.


In this work, by extending a recently introduced model capable of accounting for the disentanglement process of untensioned torus knots in 3D~\cite{Caraglio2019}, we study the mechanism of knot contour motion and knot untying for trefoil ($3_1$) and $5_1$ knots when subject to any tension.
As in Ref.~\cite{Caraglio2019}, the basic ingredients are the $n_{\mathrm{k}}$ essential crossings associated to the given knot type. 
These evolve in time as Brownian point-like particles on the rescaled support $[0,1]$ with a drift force derived from an effective free energy that depends on the knot size $\ell_{\mathrm{k}}$.
In Ref.~\cite{Caraglio2019} the free energy drives to the knot dynamics come from the excess bending energy stored in the portion of the polymer where the knot resides and the knot conformational entropy.
While the first one entails knot expansion, the latter limits it, which, in long open chains, results in a typical knot size as suggested by the recent theory of metastable knots~\cite{Grosberg2007,Dai2014,Grosberg2016}.
Here, by adding a tension dependent contribution to the free energy we can rationalize the statistical properties of the unknotting dynamics and knot decay for any value of the external force.
As representative of different qualitative behaviors, we compare our results with the strong pulling force limit and the free diffusive limit.
The results of the model are validated on those obtained by performing extensive Molecular Dynamics simulations on the full 3D problem.

The paper is organized as follows: in Section~\ref{sec_simulations} we present the Molecular Dynamics (MD) simulations used to study the problem and we delineate the basic properties of topological disentanglement physics.
In Section~\ref{sec_model} we discuss the mathematical model for the unknotting dynamics and its limiting cases.
In Section~\ref{sec_results} we benchmark the model against the MD simulations.
Finally, Section~\ref{sec_conclusions} summarizes and concludes the paper.

\section{Numerical Simulations}\label{sec_simulations}

To simulate and monitor topological disentanglement, we consider a coarse-grained model of a semi-flexible polymer with $N$ beads of diameter $\sigma$ kept together by a {\small FENE} potential and with a repulsive pairwise Lennard-Jones term providing excluded volume with hardcore range $\sigma$~\cite{Kremer1990}.
A bending potential determines a persistence length $l_p = 5\sigma$.
The chain is kept under tension $\mathcal{T}>0$ along the $z$-direction by a potential $V_{\mathcal{T}} = -\mathcal{T} (z_N - z_1 )$, where $z_1$ and $z_N$ are coordinates of the chain ends along $z$.
The system is in contact with a Langevin bath at temperature $T$ and the equations of motion are numerically solved by a velocity--Verlet algorithm and integrated through the {\footnotesize LAMMPS} simulation package~\cite{Plimpton1995}.
The characteristic simulation time is $\tau_{LJ} = \sigma \sqrt{m/k_BT}$, with $m$ the mass of each single bead.
For each knot type and value of the pulling force $\mathcal{T}$, we perform $1000$ runs with the specific knot initially strongly tied in the middle of the chain.

\begin{figure}[ht]
\includegraphics[width=1.0\columnwidth]{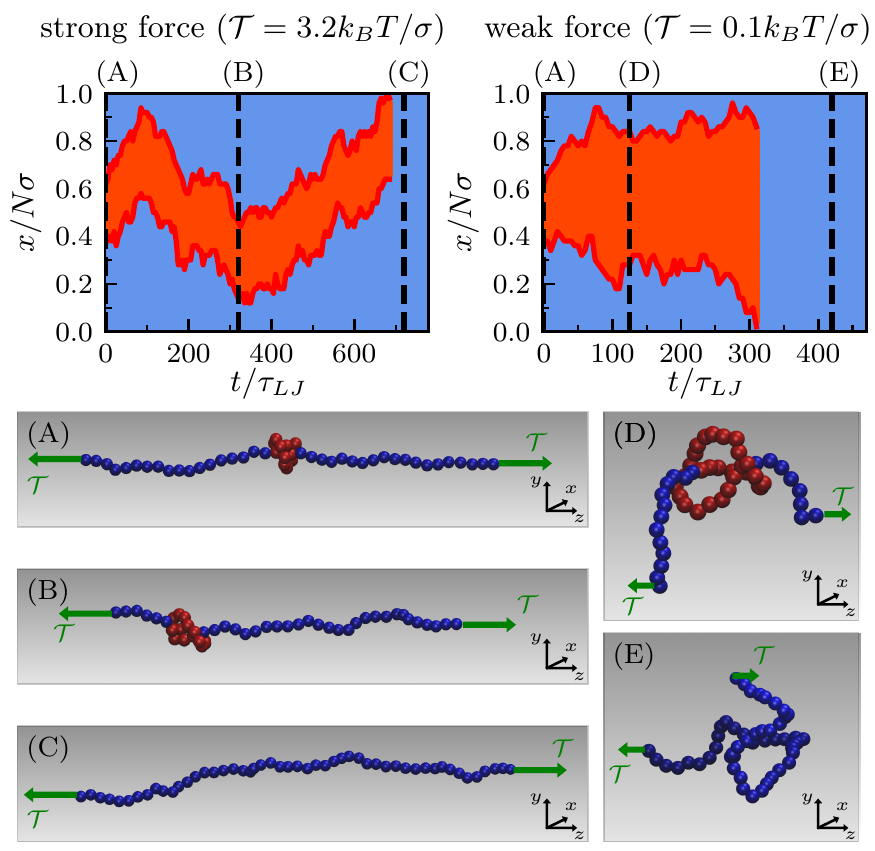}
\caption{Kymographs of the unknotting dynamics of a trefoil knot tied in a linear polymer of $N=50$ beads under strong (left) and weak (right) external pulling. 
Red highlights the knotted portion, blue indicates dangling ends; $x$ is the chemical coordinate along the backbone. 
Panels A-E display different snapshots at increasing time.}
\label{fig_model}
\end{figure}

The detection of the knotted arc along the linear chain is a crucial issue, especially at low forces, when the knot can swell and interfere topologically with the rest of the chain~\cite{Orlandini2007}.
Here we rely on an algorithm introduced in Ref.~\onlinecite{Marcone2005} and later refined and applied in several situations~\cite{Marcone2007,Tubiana2011,Tubiana2011PRL,Caraglio2015}.
For a given chain configuration all possible open portions are considered.
For each of these portions, a closure is made by joining its ends with a path specifically designed to minimally interfere topologically with the portion itself.
The knotted arc of the chain is then identified with the shortest portion still displaying the original knot type after closure (i.e. sharing the same Alexander polynomial~\cite{Adams1994}).
Through this protocol we are able to monitor the chemical coordinates along the backbone of the ends of the knot, and thus the dynamical evolution of the knot location and length. 

Typical trajectories are qualitatively different depending on the value of the pulling force.
Starting with the knot tied in the middle of the chain, when the pulling force is strong the length of the knot remains small and does not fluctuate much.
In this case, the extremities of the knot undergo a highly correlated motion so that the knot moves as a whole (left kymograph in Fig.~\ref{fig_model}).
In contrast, when the tension is sufficiently small for the polymer response to lie between the elastic and blob regimes~\cite{Matthews2010}, knot length fluctuations are important and knot ends are much less correlated (right kymograph in Fig.~\ref{fig_model}).

In this paper we limit ourselves to simple knots of the $(2,p)$-torus family~\cite{Adams1994}.
If the size of the chain is not too large or if the pulling force is strong enough, the probability of spontaneous knot formation turns out to be negligible.
Under these conditions the decay process is expected to follow the sequence $\ldots \to 7_1 \to 5_1 \to 3_1 \to\textrm{unknotted}$, i.e. a given torus knot $\mathrm{k}$ with $n_{\mathrm{k}}>3$ decays in the torus knot $\overline{\mathrm{k}}$ with $n_{\overline{\mathrm{k}}}=n_{\mathrm{k}}-2$.
In our MD simulations this picture is always confirmed.

\section{Model}\label{sec_model}
The modeling we put forward to account for the statistical properties of knot disentanglement is the one outlined in Ref.~\onlinecite{Caraglio2019}, with the addition of a tension-dependent free energy term $\Delta F_{\mathcal{T}}(\ell_{\mathrm{k}})$ (see below).
In order to be self-contained, we review here the model construction.

\begin{figure}[ht]
\includegraphics[width=1.0\columnwidth]{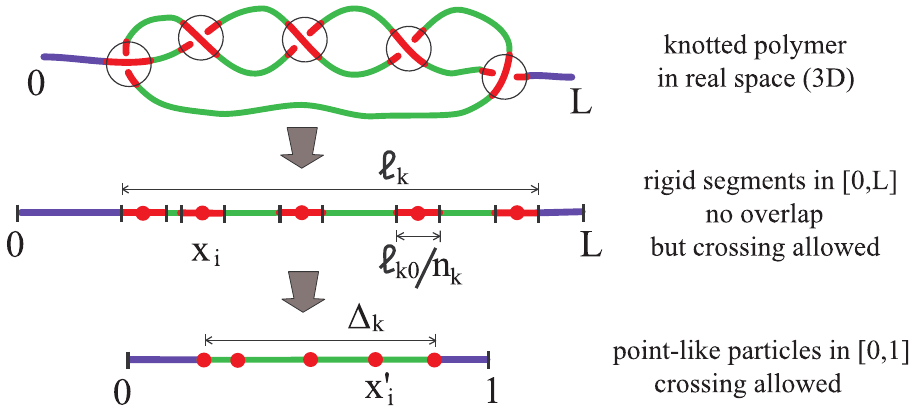}
\caption{Modeling the unknotting
  dynamics.  Central quantities are the topological invariants
  $n_{\mathrm{k}}$ (number of essential crossings) and
  $\ell_{\mathrm{k}0}/\sigma$ (knot minimal size).  A knot of contour
  length $\ell_{\mathrm{k}}$ is first mapped onto $n_{\mathrm{k}}$ rigid
  segments in $[0,L]$.  Then, rigid segments are mapped onto
  point-like particles diffusing in
  $[0,1]$. $\Delta_{\mathrm{k}}\equiv x'_{n_{\mathrm{k}}}-x'_1
  =\dfrac{\ell_{\mathrm{k}}-\ell_{\mathrm{k0}}}{L-\ell_{\mathrm{k0}}}$
  is the rescaled excess knot size.}  
\label{fig_conceptual}
\end{figure}

The main idea is to relate the motion of a knot $\mathrm{k}$ along the polymer backbone to the driven diffusion of $n_{\mathrm{k}}$ particles, $n_{\mathrm{k}}$ being the number of essential crossing.
A second key topological invariant that enters in our model is the ratio between the minimal knot length and the diameter of the chain, $\ell_{\mathrm{k}0}/\sigma$.
Here $\ell_{\mathrm{k}0}$ is the contour length of the chain portion embedding the knot when this is maximally tied.
The construction begins conceiving $n_{\mathrm{k}}$ rigid segments diffusing on the interval $[0,L]$, with $L=N\sigma$ the total contour length of the chain (see Fig.~\ref{fig_conceptual}).
Since the knot size $\ell_{\mathrm{k}}$ cannot be smaller than $\ell_{\mathrm{k}0}$, we assign the size $\ell_{\mathrm{k}0}/n_{\mathrm{k}}$ to each segment (in doing so, we implicitly take advantage of the fact that for torus knots minimal crossings are topologically equivalent) and require segments not to overlap.  
However, since for a real polymer in 3D crossings can traverse each other during the time evolution~\cite{nota1}, the segments are allowed to traverse each other.
The position along the backbone of the center of these segments is denoted by
$x_i$ ($i=1,\ldots , n_{\mathrm{k}}$) and at each time, it is convenient to renumber the segments in such a way that $\ell_{\mathrm{k}0}/2 n_{\mathrm{k}} \leq x_1 < x_2 < \ldots < x_{n_{\mathrm{k}}} \leq L-\ell_{\mathrm{k}0}/2 n_{\mathrm{k}}$.
The paradox between non-overlap and allowed-traversing of the segments is conveniently solved by a rescaling procedure: we remove the particle sizes, gluing together the remaining backbone pieces, and finally rescaling the available length $L-\ell_{\mathrm{k}0}$ to $1$ (see
Fig.~\ref{fig_conceptual}).  
Such a transformation makes the particles point-like and maps the original coordinates $x_i$ onto $x'_i \in [0,1]$:
\begin{equation}
  \label{eq_rescaling}
  x'_i \equiv \dfrac{x_i-\frac{2i-1}{2}\frac{\ell_{\mathrm{k}0}}
    {n_{\mathrm{k}}}}{L-\ell_{\mathrm{k}0}}
\qquad(i=1,2,\ldots,n_{\mathrm{k}})\; .
\end{equation}

Point-like particles satisfy overdamped Langevin equations
\begin{equation}
  \label{eq_langevin}
  \frac{\mathrm{d}x'_i}{\mathrm{d}t}
  =\alpha_i\,\frac{f'}{\zeta'}
  +\sqrt{2D'}\;\eta_i(t)
  \;,
\end{equation}
where $\eta_i$ are Gaussian white noises and $D'$ and $\zeta'$ are effective diffusion and friction coefficients, respectively.
One of our basic results is that model calibration addressed in Sec.~\ref{sec_results} suggests that $D'$ and $\zeta'$ depend on the rescaling procedure only (hence on the minimal knot length $\ell_{\mathrm{k}0}$) and are in fact independent of all other quantities like, e.g., the applied tension $\mathcal{T}$. 
The rescaled drift force $f'=f/(L-\ell_{\mathrm{k}0})$ is determined by $f=-\partial(\Delta F)/\partial{\ell_{\mathrm{k}}}$, where $\Delta F$ is the free energy difference due to the presence of the knot.
Below in this section we discuss $\Delta F$ in details, for now it is enough to anticipate that free energy drives amount to systematic expansions or contractions of the knot length and that the coefficient $\alpha_i$ homogeneously shares dilatations among the particles.
This is achieved as follows.
Since the position of the external particles is directly related to the size of the knot,
\begin{equation}
  \label{eq_knot_size}
	\ell_{\mathrm{k}} = x_{n_{\mathrm{k}}}-x_1 + \dfrac{\ell_{\mathrm{k0}}}{n_{\mathrm{k}}}
	=\ell_{\mathrm{k0}} + (x'_{n_{\mathrm{k}}}-x'_1)(L-\ell_{\mathrm{k0}}) \; ,
\end{equation}
we pose the requirements $\alpha_1\,f=-\partial(\Delta F)/\partial{x_1}$ and $\alpha_{n_{\mathrm{k}}}\,f=-\partial(\Delta F)/\partial x_{n_{\mathrm{k}}}$.
The homogeneous share of expansion or contraction is then achieved setting 
\begin{equation}
\alpha_i\equiv-1+2\dfrac{i-1}{n_{\mathrm{k}}-1}\;.
\end{equation}

\begin{figure}[ht]
\includegraphics[width=1.0\columnwidth]{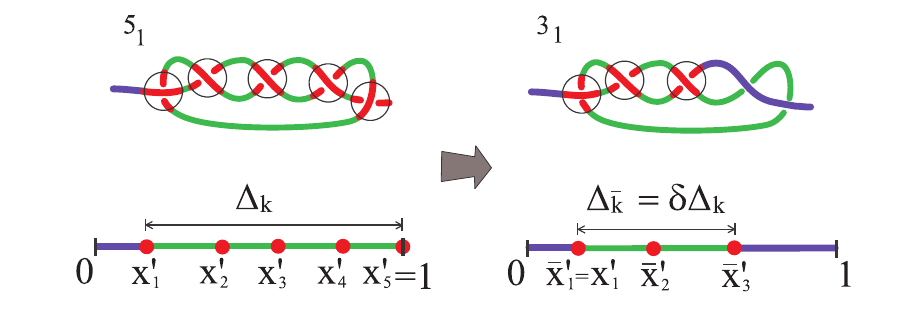}
\caption{Simplification from a $5_1$ torus knot (left) to a trefoil knot ($3_1$) (right).}
\label{fig_conceptual2}
\end{figure}

To account for knot simplification according to the sequence $\ldots \to 7_1 \to 5_1 \to 3_1 \to\textrm{unknotted}$, appropriate boundary conditions are applied. 
For a trefoil knot, whenever $x'_1 = 0$ or $x'_3 = 1$ unknotting occurs immediately.
For $n_{\mathrm{k}}>3$, whenever either $x'_1=0$ or $x'_{n_{\mathrm{k}}}=1$ the knot simplifies into a simpler torus knot $\overline{\mathrm{k}}$, with $n_{\overline{\mathrm{k}}} = n_{\mathrm{k}}-2$.
To achieve this we consistently assume that either $x'_1$ and $x'_2$ or $x'_{n_{\mathrm{k}}}$ and $x'_{n_{\mathrm{k}}-1}$ are simultaneously absorbed, respectively.
After the simplification, the positions $\overline{x}'_1, \ldots\overline{x}'_{n_{\overline{\mathrm{k}}}}$ of the point-like particles representing the crossings of the new knot $\overline{\mathrm{k}}$ are given as follows.
\begin{itemize}
  \item[(\textit{i})] If the
    $n_{\mathrm{k}}$-th ($1$-st) particle of knot $\mathrm{k}$ is absorbed, then the $1$-st ($n_{\overline{\mathrm{k}}}$ -th) particle on the new knot $\overline{\mathrm{k}}$ has the same position as the $1$-st ($n_{\mathrm{k}}$-th) particle on the old knot $\mathrm{k}$.
  \item[(\textit{ii})] The rescaled excess size of the simplified knot is a fraction $0\leq\delta<1$ of the previous rescaled excess knot size,
    \begin{equation}
      \Delta_{\overline{\mathrm{k}}}=\delta\,\Delta_{\mathrm{k}}
      \quad\Leftrightarrow\quad
      \overline{x}'_{n_{\overline{\mathrm{k}}}}-\overline{x}'_1=(x'_{n_{\mathrm{k}}}-x'_1)\,\delta\;.
    \end{equation}
Thus,
    \begin{equation}
      \overline{x}'_{n_{\overline{\mathrm{k}}}}
      = \overline{x}'_1 + \Delta_{\overline{\mathrm{k}}}
      \quad(\overline{x}'_1 =\overline{x}_{n_{\overline{\mathrm{k}}}} -
      \Delta_{\overline{\mathrm{k}}})\;,
    \end{equation}
respectively.
The other surviving particles are homogeneously placed inside the surviving knot in such a way that
    \begin{equation}
      \overline{x}'_{i}-\overline{x}'_{i-1} =
    \dfrac{\Delta_{\overline{\mathrm{k}}}}{n_{\overline{\mathrm{k}}}-1} \, , \; \forall i =
    2,\ldots,n_{\overline{\mathrm{k}}}-1\;.
    \end{equation}
Fig.~\ref{fig_conceptual2} exemplifies why the knot-size renormalization is required.
The value of $\delta$ is of course specific to the particular polymer configuration.
In our model we use the average value as measured in MD simulations when the $5_1$ knot decays into a trefoil.
\end{itemize}
After decay, also the effective diffusion and friction coefficients must be adapted to the surviving knot specifications.

We now turn to the major contributions of the knot to the polymer free energy.  
As a first approximation, the free energy difference due to the presence of the knot is the sum of three contributions depending on the knot contour length,
\begin{equation}
  \label{eq_free_energy}
  \Delta F(\ell_{\mathrm{k}}) =
  \Delta F_{\mathrm{b}}(\ell_{\mathrm{k}}) + \Delta
  F_{\mathrm{c}}(\ell_{\mathrm{k}})
  +\Delta F_{\mathcal{T}}(\ell_{\mathrm{k}})\;.
\end{equation}
The first two terms are independent of the applied tension and were already described in Ref.~\onlinecite{Caraglio2019}.
They are respectively the excess of bending energy stored in the knotted portion and the contribution to the free energy given by the conformational entropy.
Consistently with the MD simulations, the excess bending energy contribution is modeled as
\begin{equation}
  \label{eq_bending}
  \Delta F_{\mathrm{b}}(\ell_{\mathrm{k}})
  =k_{\mathrm{B}}T\left[
    \epsilon_{\mathrm{b}0}\,\frac{\ell_{\mathrm{k}0}}{\sigma}
      \,\exp\left( -
      \frac{\ell_{\mathrm{k}}-\ell_{\mathrm{k}0}}{\widetilde{\ell_{\mathrm{k}}}-\ell_{\mathrm{k}0}}
      \right)
    \right]\; ,
\end{equation}
where $\epsilon_{\mathrm{b}0}\,k_{\mathrm{B}}T$ represents the extra bending energy per monomer in the tightest knot configuration (when $\ell_{\mathrm{k}}=\ell_{\mathrm{k}0}$) and $\widetilde{\ell_{\mathrm{k}}}$ is the size of the knot beyond which the bending energy in the knotted portion of the chain relaxes to that of an unknotted chain of the same length.
From a fit to our MD, $\epsilon_{\mathrm{b}0}$ and $\widetilde{\ell_{\mathrm{k}}}$ are about independent of the knot type and on the chain length.
The excess bending energy results in a contribution to the force $f'$ in Eq.~\eqref{eq_langevin} which is positive and as such leads to a knot expansion.  The conformational entropy contribution is readily understood considering that when the dangling chains departing from the knotted region are long enough, they exert an entropic compression on the knot.
In tensionless open chains this may favor the existence of a metastable knot size~\cite{Grosberg2007,Dai2014,Grosberg2016}.
This contribution is expected to be negligible for small chains~\cite{Zheng2010,Grosberg2016} and we pragmatically approximate it with its first significant term in a Taylor expansion,
\begin{equation}
  \label{eq_conformationalentropy}
  \Delta
F_{\mathrm{c}}(\ell_{\mathrm{k}})=A_{\mathrm{c}}\,k_{\mathrm{B}}T\,\frac{\ell_{\mathrm{k}}}{\sigma}\; .
\end{equation}
Whereas in tension-free simulations we observed that $A_{\mathrm{c}} \neq 0$, only if $L \gtrsim 30\,l_p$, in the presence of a pulling force knot expansion is also hindered by the tension itself and this makes in turn the dangling ends longer.
Thus, while the conformational entropy of the knot is severely reduced, that of the longer dangling ends is less affected and, as a consequence, depending on the value of the pulling force, under tension the conformational entropy is expected to play a role also for chain shorter than $30\,l_p$.

The tension dependence in the free energy is embedded in the last term of Eq.~\eqref{eq_free_energy}, $\Delta F_{\mathcal{T}}(\ell_{\mathrm{k}})$.  
Based on MD simulations, in Appendix~\ref{sec_appA} we present a study of this term which leads to
\begin{equation}
  \label{eq_pulling_contribution}
  \Delta F_{\mathcal{T}}(\ell_{\mathrm{k}}) =\mathcal{T} (b \, \ell_{\mathrm{k}} - a_{\mathrm{k} \mathcal{T}} \, \sigma ) \; ,
\end{equation}
where $b>0$ is a parameter about independent of the pulling force and
of the knot type, whereas $a_{\mathrm{k} \mathcal{T}}$ does not
contribute to Eq.~\eqref{eq_langevin} and will therefore not be
investigated in details.  Conformed to intuition, the contribution of the
pulling potential to the force $f'$ in Eq.~\eqref{eq_langevin}, is
negative and limits then the knot expansion.

In summary, our model for the disentangle dynamics works through
Eqs.~\eqref{eq_langevin}, with appropriate boundary conditions, and
the calibrated parameters that are reported in
Section~\ref{sec_results} and encapsulated in Appendix~\ref{sec_appC}.
The model effectively reproduces the
survival probability and average disentangle time under all tension
conditions for the trefoil and $5_1$ knot.


In the next subsections we address two specific
situations in which it is possible to add insight
to Eqs.~\eqref{eq_langevin}. The first occurs when the energy drive is
dominated by tension; the second when free energy contributions
compensate each other, giving rise to free diffusion.

\subsection{Strong tension limit} \label{sec_model_SF}

With the polymer under tension, the external force contribution contrasts the knot expansion due to both the bending energy and the diffusion of the crossings.
Hence, the knot is localized and its length slightly fluctuates around a well-defined value before touching the boundaries of the chain.

Under these conditions, insight is gained focusing on the motion of the middle position only of the particles in our model,
\begin{equation}
x'\equiv\dfrac{x'_1+x'_{n_{\mathrm{k}}}}{2}\,.
\end{equation}
In order to do so, we must take into account that particles can traverse each other and that at all times they are labeled such that their coordinates are in increasing order.
Moreover, we must investigate the consequences of strong pulling in the motion of the middle position.
Appendix~\ref{sec_appB} reports a careful theoretical and numerical analysis which demonstrates that in view of these effects the middle coordinate $x'$ free diffuses with a coefficient $D'_{\text{mid}}$ which in the large tension limit is $D'_{\text{mid}} \simeq 0.33\,D'$ for the trefoil knot, and $D'_{\text{mid}} \simeq 0.21\,D'$ for the $5_1$ torus knot.

Accordingly, the probability density function (PDF) for finding the middle coordinate at $x'\in [0,1]$ at time $t>t_0$ is
\begin{equation}
\label{eq_prob}
p(x',t)
=\int_0^1 \text{d} x'_0 \; G_{D'_{\text{mid}}}(x',t|x'_0,t_0) \; p(x'_0,t_0) \; ,
\end{equation}
where $x'_0$ is the position at time $t=t_0$ and $G_{D'_{\text{mid}}}$
is the Green function of the diffusion equation $\partial p/\partial
t=D'_{\text{mid}}\;\partial^2 p/\partial x^2$ with absorbing boundary
conditions at $x'=0$ and $x'=1$: 
\begin{align} \label{eq_green}
G_{D'_{\text{mid}}} (x',t|x'_0,t_0) = &
 \;2\sum_{n=1}^\infty
 \sin (n\,\pi\,x'_0) \;\sin(n\,\pi\,x') \times \nonumber \\
 & \exp \left\lbrace -n^2\,\pi^2\,D'_{\text{mid}} \,(t-t_0) \right\rbrace \; .
\end{align}
The initial conditions in the MD simulations are always with the knot in a tight configuration in the middle of the chain.
So, at $t=t_0$ the PDF of $x'$ can be assumed to amount to a delta function $\delta(x'-x'_0)$ with $x'_0=1/2$.

The survival probability of the knot $\mathrm{k}$ is then obtained by integration:
\begin{align}\label{eq_surv}
S_{\mathrm{k}}(t)  =  \int_0^1 \text{d} x' \;  p(x',t) & =  \dfrac{4}{\pi}\sum_{n=0}^{\infty} \dfrac{\sin \left[ (2n+1) \pi x'_0 \right] }{2n+1}  \times \nonumber \\
& \exp \left\lbrace -(2n+1)^2 \pi^2 D'_{\text{mid}} t \right\rbrace  \; ,
\end{align}
and finally, the probability for the particle being absorbed at time $t$ is given by $-\partial S_{\mathrm{k}}(t) / \partial t$.

Now, suppose that the tight knot $\mathrm{k}$ is a torus knot with $n_{\mathrm{k}} > 3$ crossings.
To deal with the knot decay within this simplified context, one has to specify the coordinate $\overline{x}'$ of the middle position of the new knot $\overline{\mathrm{k}}$ arising when the coordinate $x'$ of
$\mathrm{k}$ touches one of the boundaries.
To do so we first notice that in the strong force limit $\ell_{\mathrm{k}} \simeq \ell_{\mathrm{k}0}$ and $\ell_{\overline{\mathrm{k}}} \simeq \ell_{\overline{\mathrm{k}}0}$.
This implies that $\Delta_{\overline{\mathrm{k}}} = \Delta_{\mathrm{k}}=0$ ($\delta=0$) or, equivalently, that all the rigid segments of length $\ell_{\mathrm{k}0}/n_{\mathrm{k}}$ are in contact to each other forming a single segment of length $\ell_{\overline{\mathrm{k}}0}$ (see Fig.~\ref{fig_conceptual}).
When $x'=0$ we have to remove the two segments of size $\ell_{\mathrm{k}0}/n_{\mathrm{k}}$ at the left of the knot $\mathrm{k}$ and we must take into account that the new knot $\overline{k}$ has $n_{\overline{\mathrm{k}}} = n_{\mathrm{k}}-2$ essential crossings.
The initial condition for the further diffusion of the new simplified knot is thus
\begin{equation}
\label{eq_tight_left}
\overline{x}'_0 =\frac{\frac{5}{2} \frac{\ell_{\mathrm{k},0}}{n_\mathrm{k}} -\frac{1}{2}  \frac{\ell_{\overline{\mathrm{k}},0}}{n_{\overline{\mathrm{k}}}}     }{ L-\ell_{\overline{\mathrm{k}}0} } = \dfrac{2 \ell_{\overline{\mathrm{k}},0}}{n_{\overline{\mathrm{k}}} (L-\ell_{\overline{\mathrm{k}}0})}
= \Delta_{\mathrm{k},\overline{\mathrm{k}}} \; ,
\end{equation}
where in the second equality we used the fact that the ratio $\ell_{\mathrm{k},0}/\sigma n_\mathrm{k}$ is a constant independent of the considered torus knot.
Only, the diffusion coefficient $D'_{\text{mid}}$ should be updated considering that a new knot with $n_{\overline{\mathrm{k}}}$ essential crossing is diffusing.
In a symmetric way, one argues that if the adsorption occurs at the right border ($x'=1$), then the initial position of the new knot is
$\overline{x}'_0=1-\Delta_{k,\overline{k}}$.
Eqs.~(\ref{eq_prob}, \ref{eq_green}, \ref{eq_surv}) can then be employed again with an initial distribution delta-peaked in $\overline{x}'_0$ to analyze the further decay of knot $\overline{k}$.
In general, the whole procedure can be iterated until the full disentanglement takes place.

\subsection{Free diffusion limit}

A complementary situation adding valuable insight is associated with the situation in which competing free energy effects compensate to produce an average zero drive ($f'=0$).
According to the analysis in Appendix~\ref{sec_appB} this happens for small but non-zero value of the tension $\mathcal{T}$ (see Fig.~\ref{fig_appendixB1}).

Under these conditions, different parts of the knot move independently of each
other and we can map the knot's dynamics onto that of $n_{\mathrm{k}}$
non-interacting particles which evolve in time according to
Eq.~\eqref{eq_langevin}, without the deterministic drift term.  The
initial conditions in the MD simulations are always with the knot in a
tight configuration at the middle of the chain.  So, at $t_0$ the PDFs
of the various $x'_i$ corresponds to delta functions
$\delta(x'_i-1/2)$.
At variance with the previous limit, here we must account for $n_{\mathrm{k}}$
different PDFs ($i=1, \ldots,n_{\mathrm{k}}$) of freely diffusing particles.
Correspondingly, the survival
probability of the $i$-th particle, $S_i$, can be calculated through
Eqs.~(\ref{eq_prob}, \ref{eq_green}, \ref{eq_surv}) with $D'$
replacing $D'_{\text{mid}}$ and $x'_i$ replacing $x'$.  In view of
particles' independency, the knot survival probability
$S_{\mathrm{k}}$
is now given by
\begin{equation} \label{eq_Survival_nk_particles}
S_{\mathrm{k}}(t)= \prod_{i=1}^{n_{\mathrm{k}}} S_i(t) \; ,
\end{equation}
and the probability of observing the knot decay at time $t$ becomes
\begin{equation} \label{eq_distr_decay_times}
-\dfrac{\partial S_{\mathrm{k}}(t)}{\partial t} = 
-\sum_{i=1}^{n_{\mathrm{k}}}  \left[ \prod_{j \neq i} S_j(t) \right] \, \frac{\partial S_i(t)}{\partial t} \; .
\end{equation}

At the decay time $t_1>t_0$ the marginal PDFs of the
$n_{\overline{\mathrm{k}}}$ survived particles are given by
\begin{equation}
\label{eq_prob_particles}
p(x'_{i,1},t_1)
=\int_0^1 \text{d} x'_{i,0} \; G_{D'} (x'_{i,1},t_1|x'_{i,0},t_0)\;p(x'_{i,0},t_0) \; ,
\end{equation}
where $x'_{i,0}$ denotes the position of the $i$-th particle at
$t=t_0$ and $D'$ is the diffusion coefficient associated with knot
$\mathrm{k}$. In correspondence with the length contraction introduced
by the knot's simplification, these
PDFs must be rescaled by the factor $\delta$
and can then be seen as
defining $n_{\overline{\mathrm{k}}}$ initial distributions
$p(\overline{x}_{i,1},t_1)$ that can be inserted into
Eqs.~(\ref{eq_prob}, \ref{eq_green}, \ref{eq_surv},
\ref{eq_Survival_nk_particles}, \ref{eq_distr_decay_times}) to study
the next topological simplification occurring at time $t>t_1$.  Now,
in Eq.s~(\ref{eq_prob}, \ref{eq_green}, \ref{eq_surv}) one has to
replace $x'$ with $\overline{x}'_i$, $x'_0$ with
$\overline{x}'_{i,1}$, $t_0$ with $t_1$, and $D'_{\text{mid}}$ with
$D'$, where $D'$ is the diffusion coefficient of knot
$\overline{\mathrm{k}}$.  All this procedure can be further iterated
if $\overline{\mathrm{k}}$ can still decay to simpler knots.

\section{Results}\label{sec_results}

The main quantities one can compute and compare with simulation results include the survival probability at time $t$, $S_{\mathrm{k}}(t)$, and the average survival time, $\tau_{\mathrm{k}} = \int_0^{\infty} S_{\mathrm{k}}(t) \text{d}t$, for a knot $\mathrm{k}$.
In cases in which the knot decays into a simpler one, like for an initially tied torus knot $5_1$, one can also report the probability that at time $t$ the knot is already present in the simpler form  ($3_1$ in this example) resulting from topological decay.
We perform this comparison in the case of a chain composed of $N=100$ beads.

However, before starting the comparison, it is better to provide
details about the calibration procedure; results here described are also summarized in Appendix~\ref{sec_appC}.
We start with those parameters that can be obtained independently of the point-like particle model, directly from the MD simulation data.
First, the minimal knot size $\ell_{\mathrm{k}0}$ is obtained by measuring the average size of the knot $\mathrm{k}$ when a very high force ($\mathcal{T} = 10 k_B T/\sigma$) is pulling apart the polymer ends.
This gives $\ell_{\mathrm{k}0}=13.7\sigma$ and $20.8\sigma$ for the trefoil and the $5_1$ knot, respectively.
Second, the bending energy parameters, $\epsilon_{\mathrm{b}0} \simeq 0.74$ and $\widetilde{\ell_{\mathrm{k}}} \simeq 1.85 \ell_{\mathrm{k}0}$ about independently of the knot type and of the chain length (see SI of
Ref.~\cite{Caraglio2019}).
Third, regarding the pulling energy parameters, with $N=100$ we found $b \simeq 0.6$ both for the trefoil and the $5_1$ knot about independently of the pulling force (see Appendix~\ref{sec_appA}).
Not affecting $f'$, the actual value of $a_{\mathrm{k}\mathcal{T}}$ is not reported being not relevant to our discussion.
Fourth, the parameter $\delta$ is obtained by directly measuring the size of the knot just before and after the simplification in MD simulations at null pulling force.
Specifically, we measured $\delta = 0.65$ in the case $5_1 \rightarrow 3_1$.

We next consider the parameters $D'$ and $\zeta'$.  They are obtained
as those that best fit the knots' survival probability from
independent MD simulations at null pulling force and for different
sizes of the chain ($N=50$, $100$ and $200$)~\cite{Caraglio2019}.
Remarkably, estimates suggest that $D'$ and $\zeta'$ obey
the following scaling properties with the chain size:
\begin{equation}
  D'=\frac{D_0}{(L-\ell_{\mathrm{k}0})^2}, \qquad \zeta'\;=\zeta_0\;\left( \frac{L-\ell_{\mathrm{k}0}}{\sigma}\right) ^2\; .
  \label{eq_zeta_rescaling}
\end{equation}
where $D_0$ and $\zeta_0$ are independent of the chain length $L$ and of the knot type $\mathrm{k}$.
Explicitly, we obtained $D_0=0.31$ and $\zeta_0 = 0.0016$.

\begin{figure}[ht]
\includegraphics[width=1.0\columnwidth]{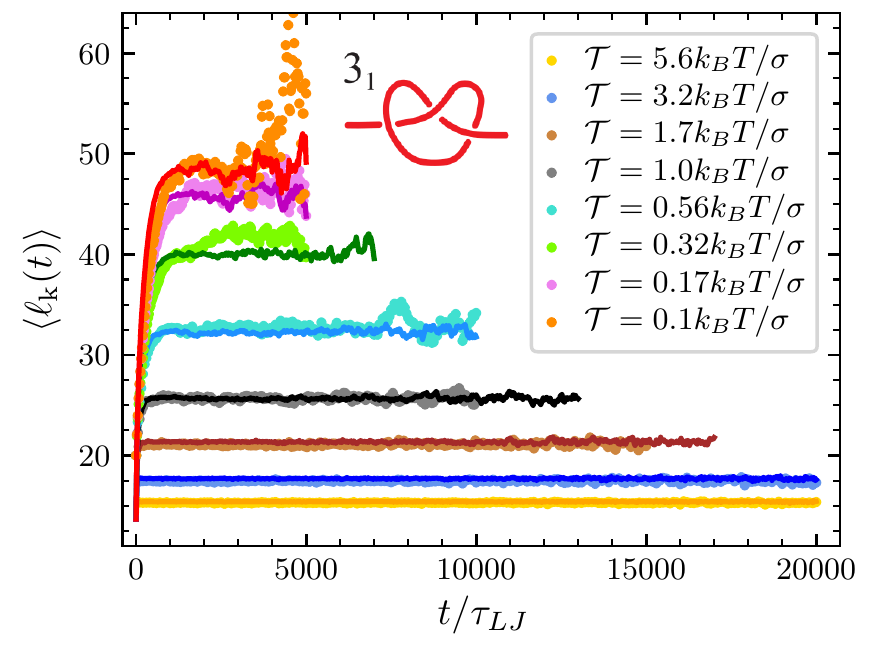}
\vspace{-20pt}
\caption{Average knot size of the $3_1$ ($n_{\mathrm{k}}=3$) at the various value of $\mathcal{T}$ considered as a function of time.
Dots are data from MD simulations and the corresponding full lines are the prediction of the model once $A_{\mathrm{c}}$ is adjusted to obtain the best agreement with MD simulations.}
\label{fig_sizetrefoil}
\end{figure}

\begin{figure}[ht]
\includegraphics[width=1.0\columnwidth]{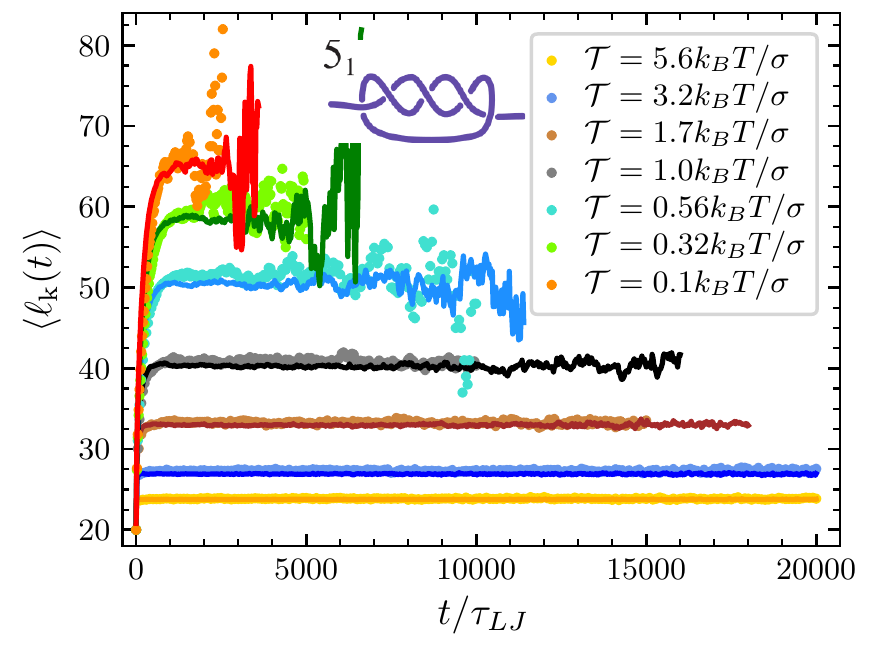}
\vspace{-20pt}
\caption{Same as Fig.~\ref{fig_sizetrefoil} but for the $5_1$ knot ($n_{\mathrm{k}}=5$).}
\label{fig_size51}
\end{figure}

Finally, the only remaining parameter that we have to estimate by directly fitting the mathematical model to the simulations at various pulling forces is the conformational entropy parameter $A_{\mathrm{c}}$. 
For each knot and each value of the pulling force considered we adjust $A_{\mathrm{c}}$ in such a way that the typical knot size predicted by our model matches the one obtained by the MD simulations (see Figs.~\ref{fig_sizetrefoil} and~\ref{fig_size51}).
In the case of the trefoil knot and $N=100$, we found $A_{\mathrm{c}} = 0.02, \, 0.05, \, 0.10, \, 0.18, \, 0.20, \, 0.24, \, 0.32$ and $1.58$ for $\mathcal{T} = 0.1, \, 0.17, \, 0.32, \, 0.56, \, 1.0, \, 1.7, \, 3.2$ and $5.6 k_BT/\sigma$ respectively.
Given the smaller size of the trefoil in comparison to the $5_1$ knot, for the latter we expect lower values of $A_{\mathrm{c}}$. Indeed for the $5_1$ knot we found $A_{\mathrm{c}} = 0.00, \, 0.02, \, 0.04, \, 0.09, \, 0.14, \, 0.11, \, 0.14$ and $1.2$ for $\mathcal{T} = 0.1, \, 0.17, \, 0.32, \, 0.56, \, 1.0, \, 1.7, \, 3.2$ and $5.6 k_BT/\sigma$ respectively.

\begin{figure}[ht]
\includegraphics[width=1.0\columnwidth]{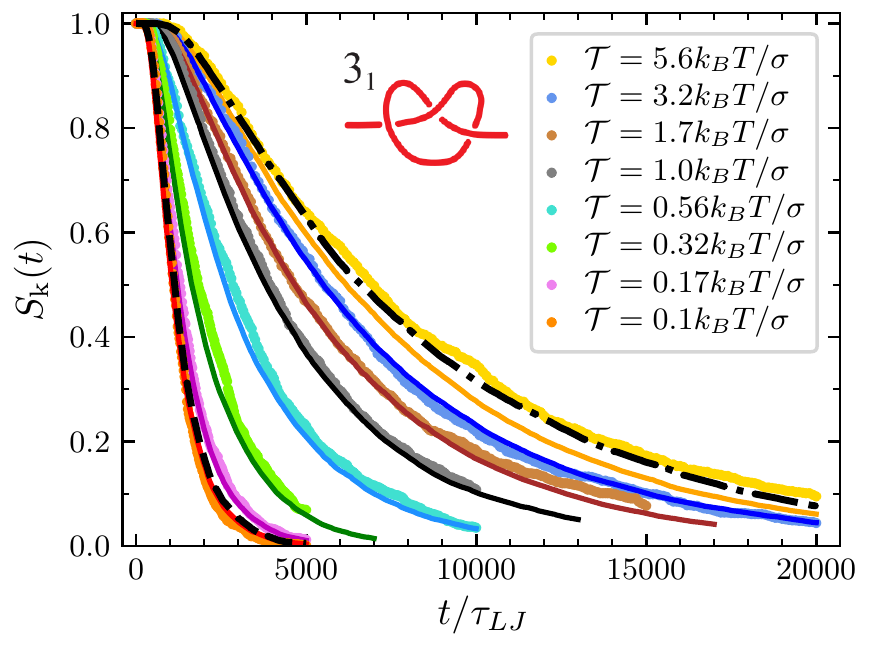}
\vspace{-20pt}
\caption{Survival probability of the $3_1$ ($n_{\mathrm{k}}=3$) at the various value of $\mathcal{T}$ considered.
Dots are data from MD simulations and the corresponding full lines are the prediction of the model.
Dashed black line on the left is the prediction of the free diffusive model with $n_{\mathrm{k}}=3$ and the same diffusion coefficient D used in the general model.
Dash-dotted line on the right is the prediction in the strong force limit.}
\label{fig_survivaltrefoil}
\end{figure}

The predictive power of our model is best illustrated by the plots of the survival probability $S_{\mathrm{k}}(t)$ reported in Fig.~\ref{fig_survivaltrefoil} for the trefoil knot.
As expected, data from MD simulations point out that the knot disentanglement process is faster when the pulling force is low, with the average survival time $\tau_{\mathrm{k}}$ increasing monotonically with the pulling force.
Except for the case of $\mathcal{T}=5.6 k_BT/\sigma$, our model nicely reproduces the behavior of data from MD simulations.
Indeed, the case of $\mathcal{T}=5.6 k_BT/\sigma$ can be better understood in the light of the strong force limit discussed in Section~\ref{sec_model_SF}.
In this limit, the model maps the knot behavior onto that of a single particle diffusing in $[0,1]$ with a diffusion coefficient $D'_{\text{mid}}$ which is about independent of the pulling force (see the plateau in Fig.~\ref{fig_appendixB1}).
The survival probability obtained in this limit is reported with a dash-dotted line reported in Fig.~\ref{fig_survivaltrefoil} and represents the limit of validity of our modeling.
While model predictions (orange line) consistently remain at the left
of this strong force limit and (we checked that this is true for any
applied tension), the survival probabilities obtained from MD simulations lay at the right of the dash-dotted line in Fig.~\ref{fig_survivaltrefoil} for $\mathcal{T}>5.6 k_BT/\sigma$.
We believe that the origin of this behavior resides in enhanced friction due to steric interaction between the various bead of an extremely compact knot, an effect which is not taken into account in our model.
On the other hand, one may add that the strong force limit model can still be used to well describe the MD simulations data also for pulling forces $\mathcal{T} \gtrsim 5.6 k_BT/\sigma$, provided that $D'_{\text{mid}}$ is directly measured by studying the mean square displacement of the knot position along the chain.

The other instructive limit is the free diffusive limit.
In this limit the free energy drives compensate and in our model the $n_{\mathrm{k}}$ particles freely diffuse with the same diffusion coefficient $D'$ used in the general case.
The trefoil survival probability in the case of pulling force $\mathcal{T}=0.1 k_BT/\sigma$ (see the lines crossing in Fig.~\ref{fig_appendixB1}) has a behavior very similar to the one described by this limit, which is represented in Fig.~\ref{fig_survivaltrefoil} by the black-dashed line.
Consistently with the fact that lines crossing in Fig.~\ref{fig_appendixB1} happens in effect slightly at the right of
$\mathcal{T}=0.1 k_BT/\sigma$, MD simulations at this value of the
tension display a slightly faster decay that what would happen in the free diffusive limit.
In such a way, we understand that at this tension the expansion effect due to the excess bending energy is not completely compensated for by the pulling and the conformational entropy drives.

\begin{figure}[ht]
\includegraphics[width=1.0\columnwidth]{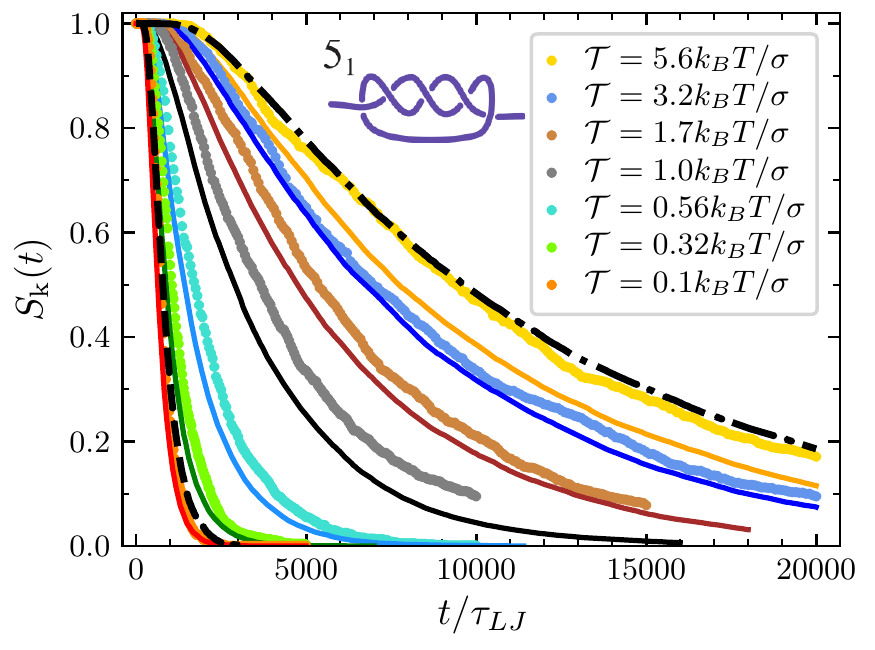}
\vspace{-20pt}
\caption{Same as Fig.~\ref{fig_survivaltrefoil} but in the case of the $5_1$ knot.}
\label{fig_survival51}
\end{figure}

The model is exploited in its full scope when we consider the full disentanglement process of more complex torus knots.
Here, we study in detail the $5_1$, i.e. the simplest torus knot after the $3_1$.
Again, survival probabilities observed in MD simulations are well reproduced for the initial decay $5_1\to3_1$ (see Fig.~\ref{fig_survival51}), and one can draw conclusions similar to those delineated above for the trefoil knot.

In comparison to the trefoil, one notes that the spread among the various curves at the different values of the pulling force is bigger.
At small forces, the $5_1$ decays faster than the trefoil while at high forces the opposite holds.
The first observation is easily understood considering that $5_1$ knot is modeled with $n_{\mathrm{k}}=5$ particles while the trefoil is modeled with $n_{\mathrm{k}}=3$ particles and that, consequently, the average time before the first particle reaches the boundary is shorter in the first case.
However, given that $D_0$ and $\zeta_0$ are independent of the knot type, one may be surprised to notice that at high forces the $5_1$ knot decays slower than the trefoil.
In fact, since the $5_1$ knot corresponds to $n_{\mathrm{k}}=5$ particles and the trefoil to $n_{\mathrm{k}}=3$ particles, one may naively expect the first one to decay faster.
The explanation of this apparent contradiction is hidden in the rescaling procedure to the unitary interval $[0,1]$.
Considering Eqs.~\eqref{eq_langevin} and~\eqref{eq_zeta_rescaling} and the fact that $f' = f/(L-\ell_{\mathrm{k}0})$, it follows that while the stochastic term is rescaled by a factor $(L-\ell_{\mathrm{k}0})$, the drift term is rescaled by $(L-\ell_{\mathrm{k}0})^3$.
Since the minimal knot size of the $5_1$ knot is greater than that of the trefoil, it follows that the drift term in the case of the $5_1$ knot has a larger weight in the rescaled dynamics than in the case of the trefoil.  And since at high pulling force the effect of the drift term is mainly contrasting the knot expansion, one can finally explain the longer lifetime of the $5_1$ knot at high forces.

\begin{figure}[ht]
\includegraphics[width=1.0\columnwidth]{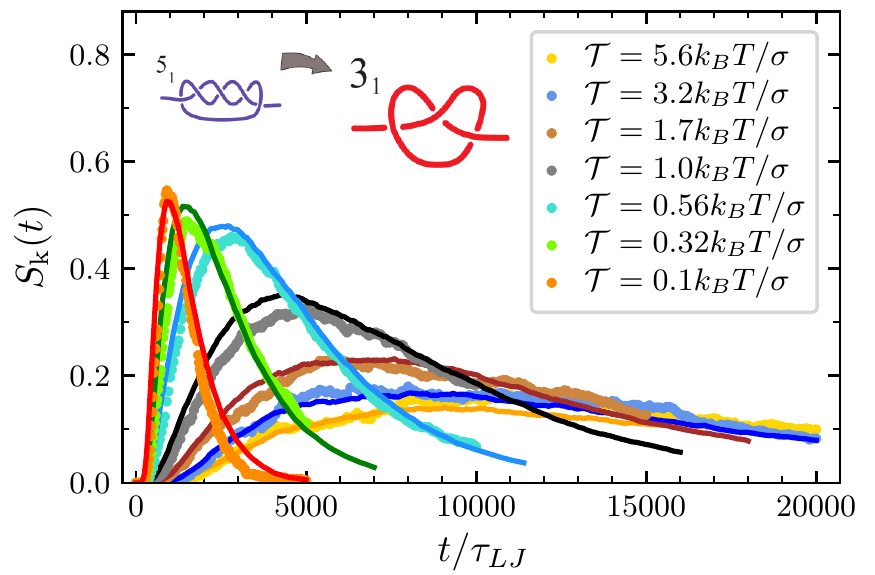}\\
\caption{Probability to have a trefoil knot after the decay $5_1 \rightarrow 3_1$.}
\label{fig_probability31}
\end{figure}

Following the procedure describe in Section~\ref{sec_model}, it is possible to study also the probability of having a trefoil after the$5_1$ simplification.
Also in this case the validity of our modeling is nicely confirmed (see Fig.~\ref{fig_probability31}).
Note that all the probabilities in Fig.s~\ref{fig_survivaltrefoil},~\ref{fig_survival51}, and~\ref{fig_probability31} die down for longer simulation times. In fact, spontaneous knot formation is not taken into account in our theoretical description and, for the considered values of $N$ and $l_p$ turns out to be negligible also in the MD simulations.
We also stress that here only chains of length $N=100$ have been considered.
However, our mathematical scheme can also be applied to other values of $N$.
In Ref.~\cite{Caraglio2019} we showed how, for $\mathcal{T}=0$, the model reproduces well the results of MD simulations also for $N=50$ and $N=200$, with $A_{\mathrm{c}}$ the only parameter dependent on the system size.
Then, provided that $A_{\mathrm{c}}$ is properly adjusted, we are confident that the present model can account for the description of the topological disentanglement occurring in chains of different length and subjected to a non-vanishing pulling force.

\section{Conclusions}\label{sec_conclusions}

Entanglement affects in a non-trivial way the mechanical properties of polymer chains subject to tensile forces~\cite{Caraglio2015,Panagiotou2014,Caraglio2017}.
In this work we have studied the untying statistics of torus knots in open chains subject to a pulling force.
The investigation is carried out both numerically via Brownian dynamics simulations and theoretically by extending a model originally introduced for free chains~\cite{Caraglio2019}.
In this model the knot dynamics is described by the diffusion of $n_{\mathrm{k}}$ particles representing the essential crossings of the underlying knot type.
Within this picture the action of the pulling force as well as the interplay between bending energy and entropy, ruling the expansion/contraction of the knot, is accounted for by adding specific free energy based potentials on the total drift force acting on these particles. 
Appropriate boundary conditions endorse the model to also describe the decay process from complex to simpler knot types within the torus knot family (knot decay).
The model of Ref.~\cite{Caraglio2019} is recovered in the limit of $\mathcal{T} \rightarrow 0$; limit that in the present case of chains composed by $N=100$ beads, is well represented by the results obtained for $\mathcal{T} = 0.1 k_BT/\sigma$.
The analytical results, once benchmarked with the corresponding Brownian simulations of a fully 3D coarse-grained model of knotted chains, show that the model nicely predicts, for a wide range of forces, both the survival probability of the initially tied knot and the probability of the new knots that may form during disentanglement.
In particular the results at intermediate forces fit well with those obtained for the previously studied cases of zero and strong force~\cite{Caraglio2019,Matthews2010, Vologodskii2006,Huang2007,Trefz2014,DiStefano2014,Narsimhan2016,Soh2019,Xu2020}. These predictions are new in literature and could be a useful guideline for future experiments of knot disentanglement of mildly elongated chains based on confinement spectroscopy techniques~\cite{Persson2009}, elongational flows~\cite{Renner2014,Narsimhan2017,Klotz2017,Soh2018}, electric fields~\cite{DiStefano2014,Klotz2018}, and stretching micro-devices~\cite{Arai1999,Bao2003}.

We believe that our model can be further exploited, modified or extended to take into account several still open issues.
For instance, here we focussed on the disentanglement of $3_1$ and $5_1$ knots as representative of the torus knot family but a systematic exploration of this process within this family can be useful considering that complex torus knots have been observed in viral capsids~\cite{Arsuaga2005,Marenduzzo2009,Marenduzzo2013}.
In contrast, to encompass other knot families like for example twist knots the model should be modified by abandoning the requirement that in the disentanglement dynamics all essential crossings are equivalent.
In fact, unlike torus knots, twist knots are characterized by the presence of a single special loop, which, if untied, it would fully disentangle the chain (unknotting number equal to $1$)~\cite{Adams1994}. 
In this respect a possible modification of the model could be the introduction of two types of particles that, once adsorbed, produce two different topological states: in one case it would lead to a transition to a simpler knot type while in the other it would disentangle the chain (direct transition to the unknotted state).
A similar approach could also be used to describe the disentanglement process in the  more subtle case of slip-knots~\cite{King2007,Millett2010}

Finally, we note that in the present approach, we do not consider the possible reappearance of a knot~\cite{Tubiana2013} during the disentanglement process, a phenomenon that however becomes more and more probable as the contour length of the polymer increases and the tension is sufficiently small.
Yet, a further extension of the model, specifying for instance rules for the creation of new particles, could account also for this aspect.

\appendix

\section{Tension dependence in the free energy} \label{sec_appA}
Since our simulation protocol works at constant tension
$\mathcal{T}>0$, the polymer free energy is characterized by a term
\begin{equation}
F_{\mathcal{T}} = - \mathcal{T}  L_z \; ,
\end{equation}
where $L_z$ is the component along the pulling axis $z$ of the polymer end-to-end length.  
For unknotted polymer chains, $L_z$ has a typical average value which depends on the intensity of the pulling force $\mathcal{T}$.
Correspondingly, the free energy takes a reference value $F_{\mathcal{T}0}$.

\begin{figure}[ht]
\includegraphics[width=1.0\columnwidth]{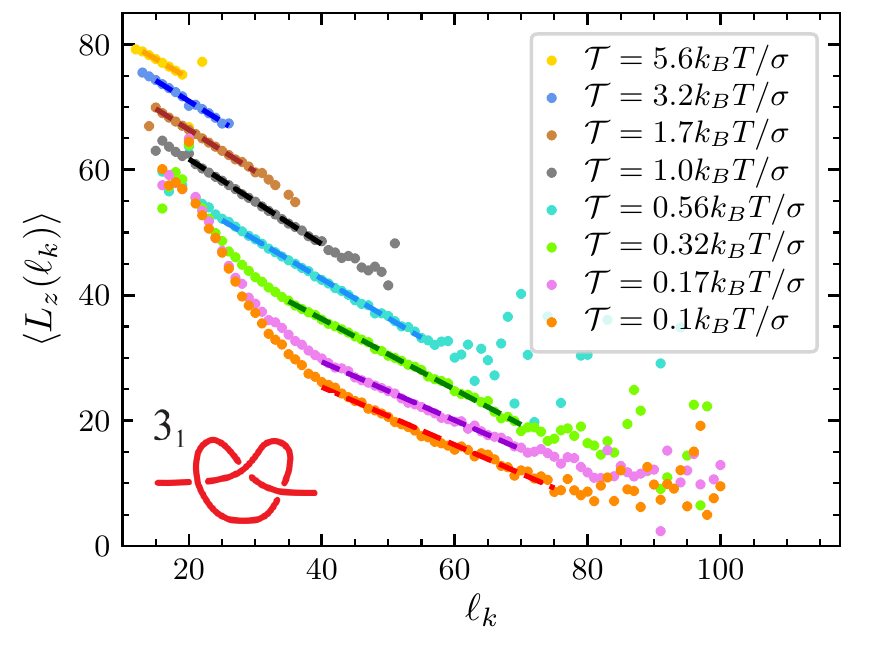}
\vspace{-20pt}
\caption{Average component of the end-to-end polymer extension along
  the direction of the pulling force as a function of the knot size,
  for the $3_1$ knot. Various value of $\mathcal{T}$ are considered.
  Dots are data from MD simulations starting with a trefoil knot tied
  in the middle of a chain of $N=100$ beads.  Dashed lines are fit to
  Eq.~\eqref{eq_pulling_contribution} in the ranges of
  $\ell_{\mathrm{k}}$ spanned by the dashed lines themselves.  Fits
  return $b_{\mathrm{k}} \simeq 0.6$ about independently of the
  pulling force and knot type.}
\label{fig_appendixA1}
\end{figure}

\begin{figure}[ht]
\includegraphics[width=1.0\columnwidth]{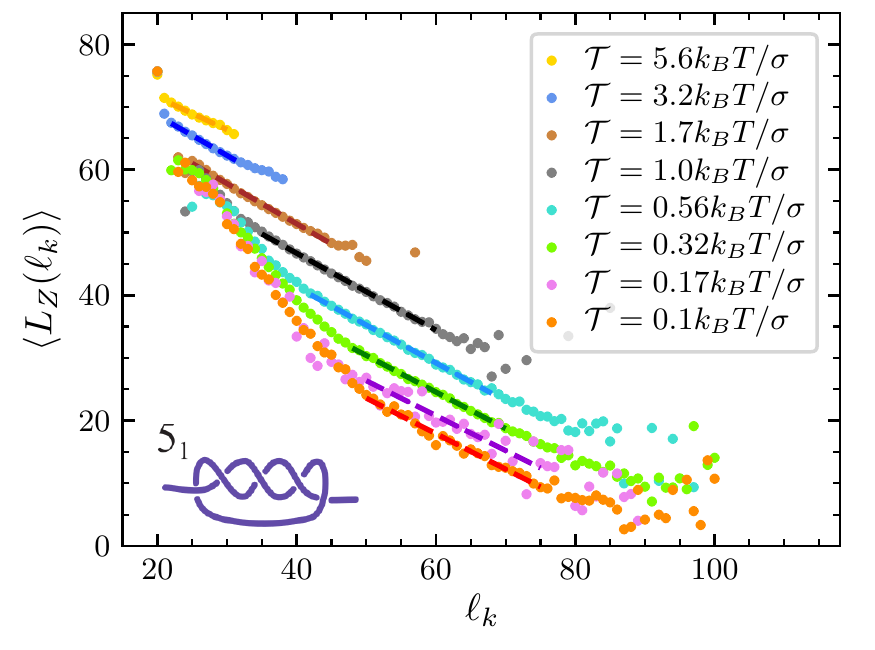}
\vspace{-20pt}
\caption{Same as in Fig.~\ref{fig_appendixA1} but in the case of the
  $5_1$ knot.}
\label{fig_appendixA2}
\end{figure}

In the presence of the knot, $\langle L_z\rangle$ becomes of course a function of the knot length, $\langle L_z\rangle=\langle L_z(\ell_{\mathrm{k}})\rangle$.  Figures~\ref{fig_appendixA1} and~\ref{fig_appendixA2} display such a dependence in the case of the trefoil and of the $5_1$ knot, respectively.  
Averages are performed using the whole set of data at disposal from our MD simulations.
Inspection of the results in the strong force regime ($\mathcal{T} = 5.6, \, 3.2 \mbox{, and } 1.7k_BT$) suggest the relation
\begin{equation}
  \langle L_z (\ell_{\mathrm{k}}) \rangle
  = a_{\mathrm{k} \mathcal{T}} \,\sigma - b\, \ell_{\mathrm{k}} \; 
\end{equation}
to hold, where $a_{\mathrm{k} \mathcal{T}}$ is a parameter that
depends both on the knot type and on the applied force, while $b$ is a
parameter that is about $0.6$ independently of both the knot type and
of $\mathcal{T}$.  Interestingly, the same relation also holds for
weak applied forces, but only for large enough values of
$\ell_{\mathrm{k}}$.  In effect, due that we start simulations with
an initial configuration in which the knot is strongly tied
in the center of a very stretched chain (see panel
A in Fig.~\ref{fig_model}), when $\ell_{\mathrm{k}}$ is small results
are strongly biased by the initial knot expansion due to the release
of the bending energy.  As such, they should not be considered in the
construction of the tension dependence in the free energy.  In
summary, our results suggest
\begin{equation}
  \Delta F_{\mathcal{T}}(\ell_{\mathrm{k}})
  =F_{\mathcal{T}}(\ell_{\mathrm{k}})-F_{\mathcal{T}0}
  =- \mathcal{T}\,\left(a_{\mathrm{k} \mathcal{T}} \,\sigma - b\, \ell_{\mathrm{k}}\right)\;. 
\end{equation}

\section{Diffusion coefficient for the middle position of $n$ trespassing particles} \label{sec_appB}
  
We address the problem of finding the diffusion
coefficient $D'_{\text{mid}}$ of the middle coordinate
$x' = (x'_1 +x'_{n})/2$ in a system of $n$ independent particles diffusing
according to the model of Section~\ref{sec_model}.

We start by considering a free diffusion ($f'=0$) in a one-dimensional
infinite support.  Let us suppose to
have $n$ independent particles free diffusing with diffusion
coefficient $D'$, and let us
denote with $-\infty<x'_1(t) \le x'_2(t) \le \ldots \le x'_n(t)<\infty$ their
position at time $t$.  At different time, the same index $i$ may refer
to different particles since the latter are allowed to cross each
other.  For simplicity we assume the initial condition $ x'_1(0) =
x'_2(0) = \ldots = x'_n(0) = 0$.  We are interested in finding a
relation between the diffusion coefficient $D'_{\text{mid}}$ of the
coordinate $x \equiv (x'_1 + x'_{n})/2$ and the diffusion coefficient
$D'$ of the single particles.  We are about to show that
in this case 
\begin{equation}
D'_{\text{mid}} = g_n \, D'  \;,
\end{equation}
with $g_n$ an $n$-dependent coefficient.
By definition, we have
\begin{equation} \label{eq_MSDeff}
\! \! \left\langle  \left( \dfrac{x'_1(t)+x'_n(t)}{2} \right)^2 \right\rangle  - \left\langle  \dfrac{x'_1(t)+x'_n(t)}{2}  \right\rangle ^2 = 2 D'_{\text{mid}} t \; .
\end{equation}
Because of the symmetry $\langle x'_1(t) \rangle = - \langle x'_n(t)
\rangle $, the previous equation becomes
\begin{equation}
\langle (x'_1)^2(t) \rangle + \langle (x'_n)^2(t) \rangle + 2 \langle x'_1(t) \cdot x'_n(t) \rangle = 8 D'_{\text{mid}} t \; ,
\end{equation}
where

\begin{widetext}
  \begin{eqnarray}
    \label{eq_integral_1}
\langle (x'_1)^2(t) \rangle &=& 
n! \int_{-\infty}^{+\infty} \text{d}y_1 \, y_1^2 \, \dfrac{e^{-\frac{y_1^2}{4Dt}}}{\sqrt{4\pi D t}} \, \int_{y_1}^{+\infty} \text{d}y_2 \, \dfrac{e^{-\frac{y_2^2}{4Dt}}}{\sqrt{4\pi D t}} \, \ldots \, \int_{y_{n-1}}^{+\infty} \text{d}y_n \, \dfrac{e^{-\frac{y_n^2}{4Dt}}}{\sqrt{4\pi D t}}
\; ;\\
\label{eq_integral_2}
\langle (x'_n)^2(t) \rangle &=& 
n! \int_{-\infty}^{+\infty} \text{d}y_1 \, \dfrac{e^{-\frac{y_1^2}{4Dt}}}{\sqrt{4\pi D t}} \, \int_{y_1}^{+\infty} \text{d}y_2 \, \dfrac{e^{-\frac{y_2^2}{4Dt}}}{\sqrt{4\pi D t}} \, \ldots \, \int_{y_{n-1}}^{+\infty} \text{d}y_n \, y_n^2 \, \dfrac{e^{-\frac{y_n^2}{4Dt}}}{\sqrt{4\pi D t}}
\; ;\\
\label{eq_integral_3}
\langle x'_1(t) \cdot x'_n(t) \rangle &=& 
n! \int_{-\infty}^{+\infty} \text{d}y_1 \, \dfrac{e^{-\frac{y_1^2}{4Dt}}}{\sqrt{4\pi D t}} \, \int_{y_1}^{+\infty} \text{d}y_2 \, \dfrac{e^{-\frac{y_2^2}{4Dt}}}{\sqrt{4\pi D t}} \, \ldots \, \int_{y_{n-1}}^{+\infty} \text{d}y_n \, (y_1 \cdot y_n) \, \dfrac{e^{-\frac{y_n^2}{4Dt}}}{\sqrt{4\pi D t}}
\; .
\end{eqnarray}
\end{widetext}
For $n=2$, it is easy to check
\begin{eqnarray}
\langle (x'_1)^2(t) \rangle = \langle (x'_n)^2(t) \rangle & = &  2 D' t \, , \\
\langle x'_1(t) \cdot x'_n(t) \rangle & = & 0 \, ,
\end{eqnarray}
which implies $D' = 2D'_{\text{mid}}$, i.e. $g_2 = 1/2$.
In a similar way, with $n=3$ 
\begin{eqnarray}
\langle (x'_1)^2(t) \rangle = \langle (x'_n)^2(t) \rangle & = & \dfrac{(\sqrt{3} + 2\pi)D' t}{\pi} \, , \\
\langle x'_1(t) \cdot x'_n(t) \rangle & = & - \dfrac{2\sqrt{3}}{\pi} D' t \, ,
\end{eqnarray}
implying $D' = 4 \pi D'_{\text{mid}}/(2\pi - \sqrt{3})$ and
$g_3\simeq 0.362$. Numerical integration of 
Eqs.~(\ref{eq_integral_1}, \ref{eq_integral_2}, \ref{eq_integral_3})
provides $g_5\simeq 0.252$ and $g_7\simeq 0.232$.

\begin{figure}[ht]
\includegraphics[width=1.0\columnwidth]{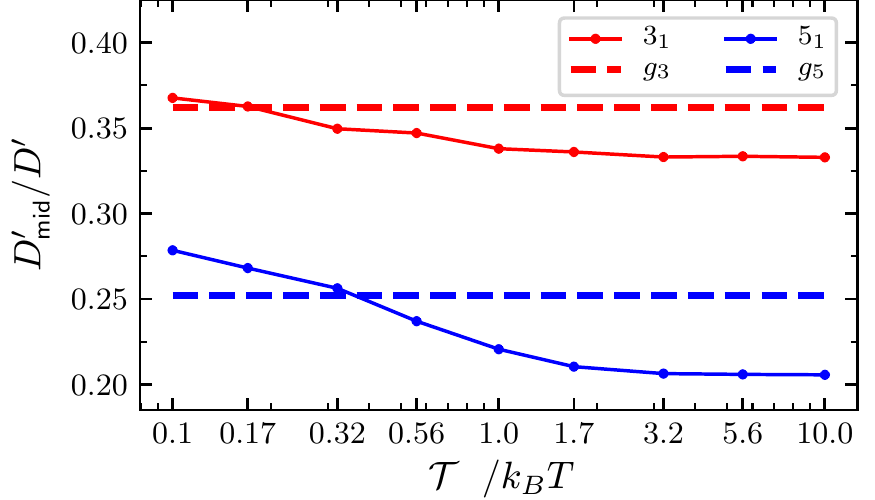}
\vspace{-20pt}
\caption{Analysis of the diffusion coefficient $D'_{\text{mid}}$ for the middle
  position of $n_{\mathrm{k}}$ particles satisfying
  Eqs.~\eqref{eq_langevin} on an infinite support. Full lines:
  $g_n=D'_{\text{mid}}/D'$ as a
  function of the pulling force $\mathcal{T}$ with
  $n=n_{\mathrm{k}}=3$ for the trefoil and $n=n_{\mathrm{k}}=5$  for the
  $5_1$ knot. Dashed lines represent the corresponding value in the
  free diffusive case ($f'=0$).}
\label{fig_appendixB1}
\end{figure}

The drift term in
Eq.~\eqref{eq_langevin} acts in such a way that with $f'>0$ 
the knot inflates and with $f'<0$ it squeezes.
This amounts to push further and closer, respectively, the first and
the $n=n_{\mathrm{k}}$-th particles compared to free diffusion.
The overall effect of such a driving
introduces a systematic correction to the above values $D'_{\text{mid}}$,
which we evaluate numerically.
Starting with the initial condition $x'_i(0)=0$, we
integrated Eq.~\eqref{eq_langevin} on an infinite support (without
boundary conditions) up
to $t=1000$ iteration steps.
Fig.~\ref{fig_appendixB1} reports the values of
$D'_{\text{mid}}$ obtained by fitting Eq.~\eqref{eq_MSDeff} to the
numerical integration data.  With low pulling force, the excess
bending energy dominates the drift term making it expansive.
Correspondingly, $D'_{\text{mid}}$ is large. As the expansive drive
turns into contractive with increasing tension, $D'_{\text{mid}}$
monotonically decreases crossing the free diffusive case, and
tends to a plateau for strong forces. 
Interestingly, 
in the case of the $5_1$ knot, $D'_{\text{mid}}$ crosses the 
free diffusive value for a larger tension $\mathcal{T}$ than the 
trefoil knot.  This is because the excess bending energy gives a
larger contribution in the case of the $5_1$ knot than for trefoil.
The asymptotic values of the effective diffusion
coefficient are about $D'_{\text{mid}} = 0.333 D'$ for the trefoil and
$D'_{\text{mid}} = 0.206 D'$ for the $5_1$ knot.  These are the values
used in Figs~\ref{fig_survivaltrefoil} and~\ref{fig_survival51} to
obtain the strong force limit (dash-dotted line).

\section{Summary of model parameters} \label{sec_appC}

Here we sum up the parameters entering in the model described in section~\ref{sec_model}, with some comments about their meaning.

\subsubsection*{Parameters independent of the knot type and of the pulling force:}

\begin{itemize}
\item $\epsilon_{\mathrm{b}0}$ represents the extra bending energy (in unit $k_BT$) per monomer in the tightest knot configuration with respect to the equipartition value $k_{\mathrm{B}}T$ attained for the corresponding relaxed state $\ell_{\mathrm{k}}\gg\widetilde{\ell_{\mathrm{k}}}$.
$$\epsilon_{\mathrm{b}0} \simeq 0.74$$
\item $D_0 = D' (L-\ell_{\mathrm{k}0})^2$, where $D'$ is the diffusion coefficient of the point-like particles representing essential crossings moving on the unit segment $(0,1)$.
$$ D_0 = 0.31 $$
\item $\zeta_0 = \zeta' / \left( \frac{L-\ell_{\mathrm{k}0}}{\sigma} \right) ^2$, where $\zeta'$ is the friction constant to be used in the Langevin equation describing the dynamics of the point-like particles.
$$ \zeta_0 = 0.0016$$
\item $b$ is the proportionality constant linking the average end-to-end length of the polymer chain to the size of the knot embedded in the chain.
$$ b = 0.6 $$
\end{itemize}

\subsubsection*{Parameters dependent on the knot type and independent of the pulling force:}

\begin{itemize}
\item $\ell_{\mathrm{k}0}$ is the minimal length of the knotted arc.
\begin{center}
\begin{tabular}{c|c}
   $\mathrm{k}$   &  $\ell_{\mathrm{k}0}$ \\
\hline
$3_1$ & $13.7 \sigma$ \\
$5_1$ & $20.8 \sigma$ \\
\end{tabular}
\end{center}
\item $\widetilde{\ell}_{\mathrm{k}}$ is the typical characteristic length in the decaying of the bending energy of the knotted portion.
$$ \widetilde{\ell}_{\mathrm{k}} = 1.85 \ell_{\mathrm{k}0}$$
\item $\delta$ is the parameter describing the reduction of the knot size in the rescaled support $[0,1]$ when a torus knot $\mathrm{k}$ decays into a simpler torus $\bar{\mathrm{k}}$.
$$ \delta = 0.65 \qquad \mbox{for} \quad 5_1 \rightarrow 3_1 $$
\end{itemize}

\subsubsection*{Parameters dependent on the knot type and on the pulling force:}

\begin{itemize}
\item $A_{\mathrm{c}}$ is the parameter measuring the contribution of the entropic force.
\begin{center}
\begin{tabular}{c|cccccccc}
   \backslashbox{$\mathrm{k}$}{$\mathcal{T}$} &  $0.1$ & $0.17$ &  $0.32$ & $0.56$ & $1.0$ & $1.7$ & $3.2$ & $5.6$\\
\hline
$3_1$ & $0.02$ & $0.05$  & $0.1$ & $0.18$  & $0.2$ & $0.24$   & $0.32$ & $1.58$\\
$5_1$ & $0$ & $0.02$  & $0.04$ & $0.09$  & $0.14$ & $0.11$ & $0.14$ & $1.2$\\
\end{tabular}
\end{center}
\end{itemize}

{\bf Acknowledgments}
MC acknowledges financial support from the Austrian Science Fund (FWF): P 28687-N27.


\begin{thebibliography}{58}
\bibitem [{not()}]{nota1}%
  \BibitemOpen
  \href@noop {} {\bibinfo  {journal} {The concept of a crossing in 3D is not
  well defined but a fluctuating polymer can be projected onto an arbitrary
  plane. On this, the dynamics may give rise to Reidemeister moves in which
  crossings can be created or annihilated and can traverse each other.}\
  }\BibitemShut {NoStop}%
\end{thebibliography}
\end{document}